\documentstyle[11pt,newpasp,twoside,epsf,psfig]{article}

\begin{document}

\title{Simulations Applied to the Bright SHARC XCLF: Results and Implications}
\author{M. P. Ulmer, C. Adami, R. Pildis}\affil{Northwestern University, (+LAS
C. Adami)}

\author{A. K. Romer, R. C. Nichol}\affil{Carnegie Mellon University}

\author{B. Holden}\affil{University of Chicago}

\begin{abstract} We have performed simulations of the effectiveness of the
Serendipitous High-redshift Archival ROSAT Cluster (SHARC) survey for various
model universes. We find, in agreement with work based on a preliminary set of
simulations (Nichol et al., 1999), no statistically significant evolution of the
luminosity function out to z = 0.8. 
\end{abstract}


\section{Introduction}

The interest in the X-ray Cluster Luminosity Function (XCLF) has been heightened in
recent years with the realization that a measure of its evolution with redshift
could be used to derive
the value of the cosmological parameter $\Omega_o$ (cf. Nichol et al., 1999,
Oukbir $\&$ Blanchard, 1997).   The conclusions drawn from this type of analysis
are, however, extremely model dependent, and observational difficulties aside, it
may be a long time before a consensus is reached as to the value of $\Omega_o$
(cf. Reichart et al. 1999, Bahcall et al. 1997).
Simply understanding the origin and evolution of clusters of galaxies is interesting
itself, however, and a measure of the evolution of the XCLF provide insights into
the process of cluster formation.  As we will describe below, it is interesting
to note that, although there is no statistically significant evidence in our 
Bright SHARC
survey for evolution of the XCLF out to a z =
0.8,  there is a great deal of evidence that the X-ray emission (and other
properties of clusters) are continually evolving.  Therefore, some how clusters
evolve, but manage to keep the total luminosity function relatively constant.  
The process that could do this is hierarchical clustering (cf. Doroshkevich et al. 1998)
in which smaller X-ray fainter clusters evolve into larger, X-ray brighter clusters, in just 
such a way so as to keep the XCLF approximately constant with redshift. 
The conclusion then is that the most likely portion of the XCLF to detect 
evolution is at the bright 
end where the
hierarchical coalescence requires the most amount of time to form such large
structures.  Below, we begin by giving just
one example of cluster evolution.  Then we describe the results of our 
simulations
and the impact of the SHARC derivation of the luminosity function for 
$ 0.3 \la z \la 0.8$. Next, we discuss how these assumptions that are included
in the simulations
affect the conclusions that can be drawn about evolution. Finally, we
demonstrate that not only is better sky coverage needed to improve statistics, 
but
the evolution of the {\em shape} of the emission  profiles is also going to be
necessary before the true completeness of a survey can be accurately determined.   

\section{Cluster Evolution}

There are many pieces of evidence that clusters are continually forming and
evolving, but one of the earliest was the 
surprising 
discovery that the
intra-cluster medium in the Perseus cluster is cooler in the core region (Ulmer
\& Jernigan, 1978). 
In fact, one of the explanations that Ulmer and Jernigan gave for this result 
was that
the denser material in the core had cooled faster due to the cooling time 
being related to
the density of the material.  This {\em first } discovery is often ignored, 
yet
this was the discovery of what now seems to be the nearly ubiquitous 
phenomenon called ``cooling flows'' (cf. Peres {\em al.}, 1998).

\section{The Simulations}

A key to any survey is to understand how effective the survey was at detecting
objects.  Without this understanding, the detected number of objects 
cannot be converted into a ``true'' number. 
The so-called ``Bright SHARC'' survey and been described in detail by Romer 
et al. (1999). 
A preliminary measure of the XCLF derived from the results of 
Romer et al. was 
reported by Nichol et al.(1999). Full
details of our simulations can be found in Adami et al.(1999).

\subsection{The Simulation Procedure}
The simulation
program produced fake clusters which were placed on real ROSAT data sets that
were used for analysis.  The fake clusters were systematically placed
within different annuli (but randomly other wise) in the ROSAT PSPC field of
view.  The data set we used was a statistically complete sampling of the data 
set that actually was used for analysis and which included all ROSAT pointing 
with galactic lat. greater than $|20^{\circ} |$ and data within the region
between $2\farcm5$ and $19'$ of the center of the field of view (FOV) of
the ROSAT PSPC. These data sets were then put through the standard SHARC
processing pipeline and extended sources were identified in the standard
manner. If a fake cluster was found as an extended X-ray source, then the 
cluster was
counted as ``detected.'' We also kept track of the apparent luminosity of the
cluster to determine how much the cluster luminosity differed from the true
(input) luminosity.

\begin{figure}[bht]
\vspace*{-1.in}
\hspace*{0.4 in}\plottwo{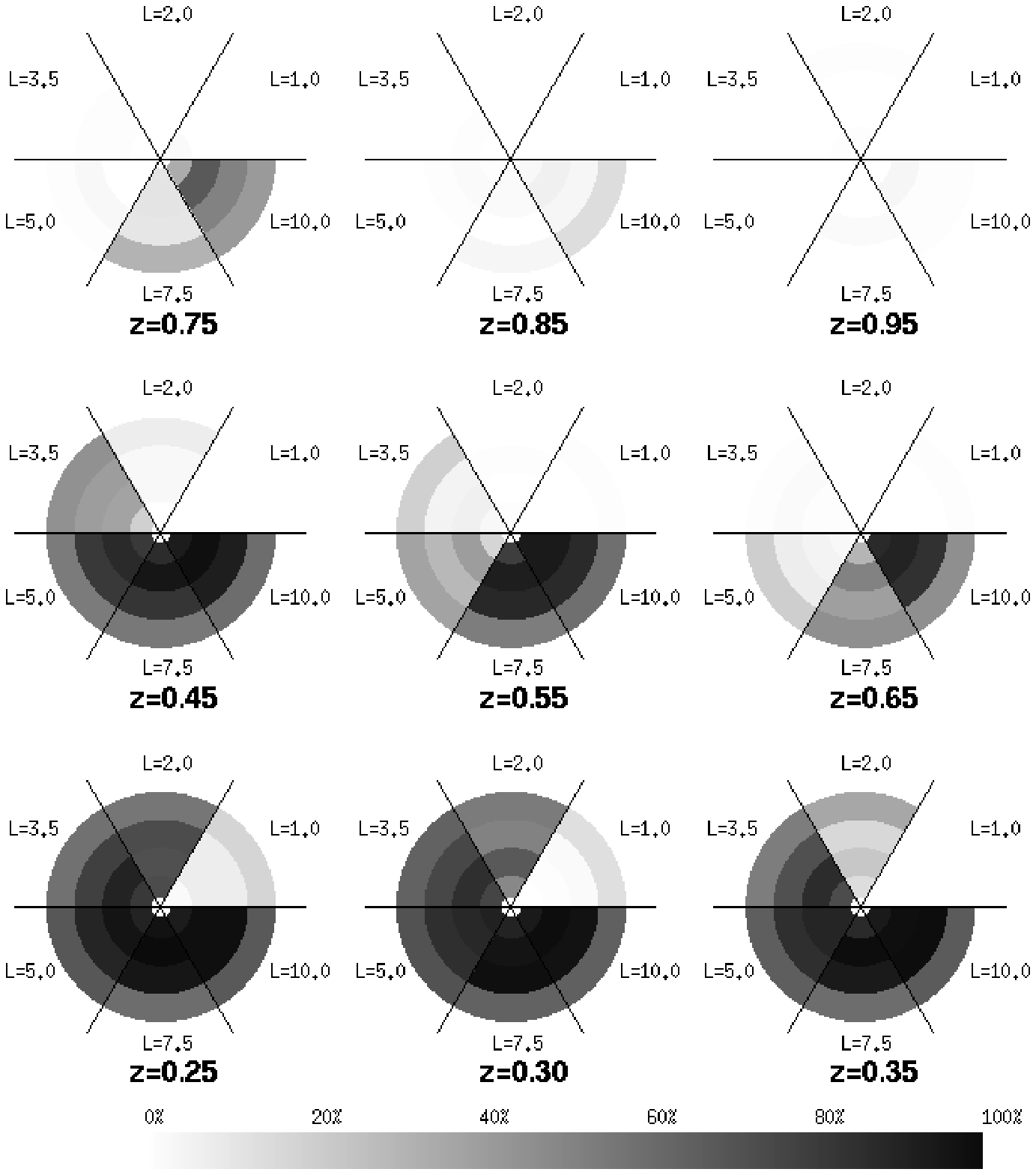}{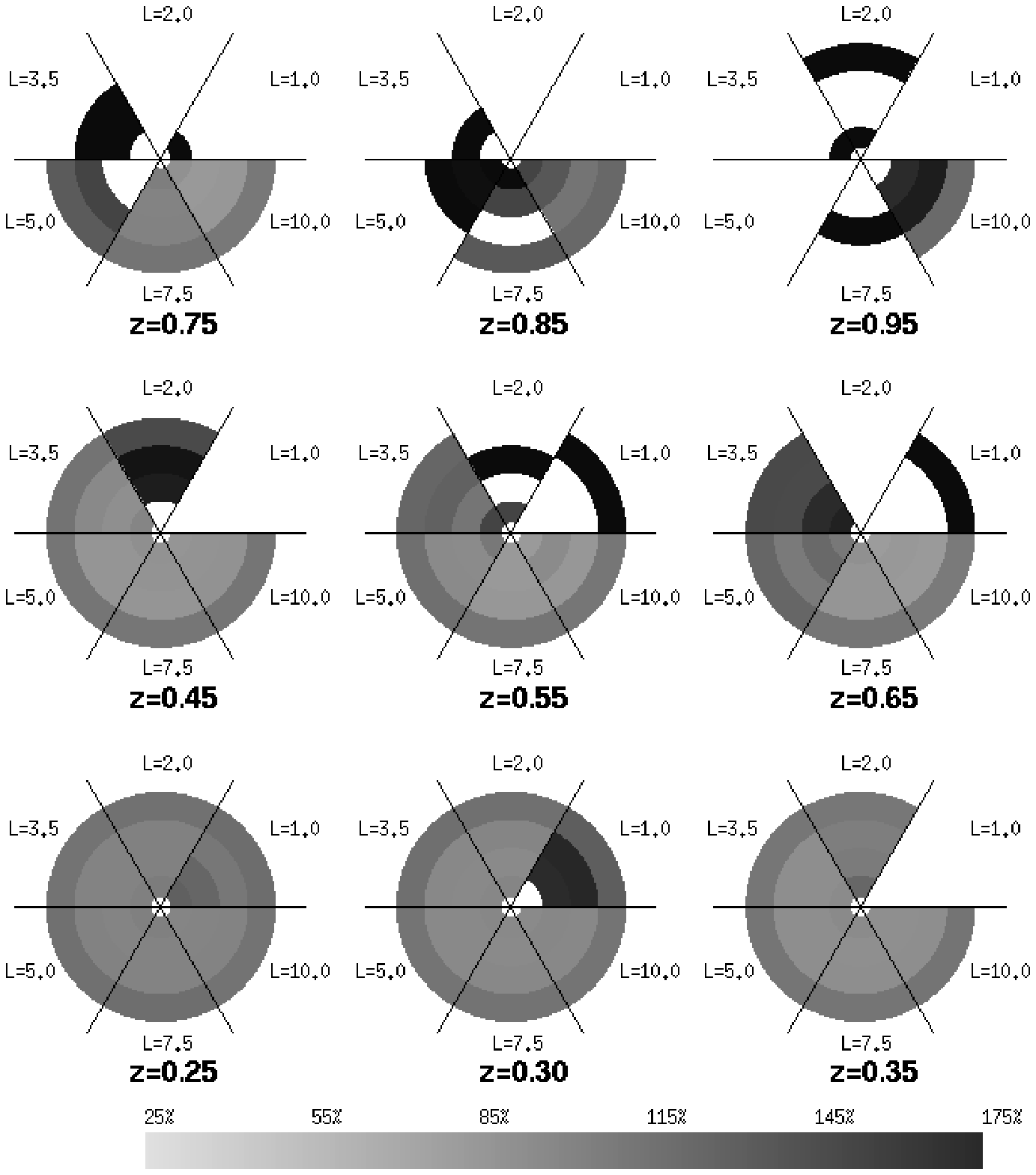}
\vspace{0.4in}
\caption{Left: Efficiency for the standard set
of parameters. Each disk represents a ROSAT PSPC FOV with the 4 tested 
different annuli. We have split the disks in 6 angular sub-sections to 
represent the results for the 6 different luminosities (L = 1, 2, 3.5, 5, 7.5, 
and 10 $\times$ 10$^{44}$ erg.s$^{-1}$). The gray scale levels are produced in
percentage at the bottom of the figure. See Adami et al. for more 
details. Right: same as the left side except the gray scale levels are the 
percentage of the true luminosity which is recovered for the Bright SHARC 
clusters. \label{fig2}}
\end{figure}
\vspace*{- .1in}

\subsection{The Results}
In Figure 1, we show two results. In the left-most panel, we show the detection
efficiency as a function of redshift and radial distance from the center of the
ROSAT PSPC FOV.  We can see the effects of the degradation of the
angular resolution in that clusters are less easily found in the outer portions
of the FOV. We also see that the detection efficiency falls off with increasing
redshift, as expected.  At redshifts $\ga 0.55$, we see some apparent
detections at low luminosity, but a comparison with the right most figure reveals the
cause: confusion. The right-most panel shows that the apparent luminosity of 
these
objects is much higher than the true luminosity which can only mean that the fake
cluster in our simulation landed close enough to a real object to have been
mistakenly classified as a ``detection.''   This turns out to be negligible 
effect,
however, as the overall detection efficiency rate is small (usually less that
5\%). This shows,
however, the importance of optical follow up and rigorous identification before
an ``extended source'' can be truly identified as an X-ray bright cluster.
We also see in Figure 1 that, generally, the luminosity was well determined by 
our
process except in cases where the detection rate is so small that in real life
(as opposed to our simulations) they were
easily removed from our sample by optical followup.  These results are all for a
standard set of parameters for the cluster profile, temperature, and cosmology
(see Nichol et al. 1999).  

\begin{figure}[bht]
\vspace{-2.in}
\hspace*{.1in}\centerline{\psfig{file={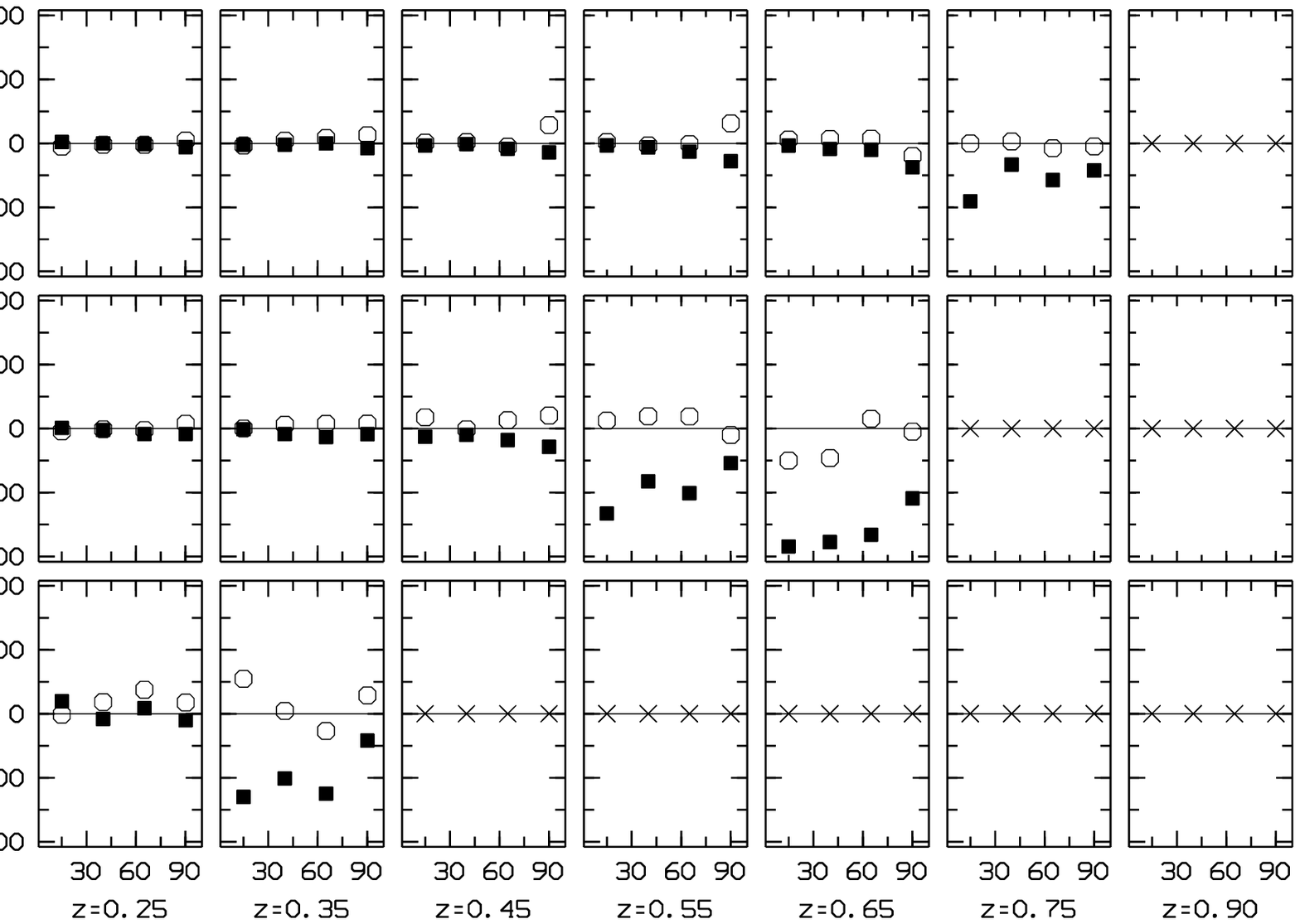},height=4.in}}
\caption{Shown is the ratio in percent bewteen the difference of the 
Bright SHARC detection efficiency with
standard parameters and detection efficiency with either a cooling flow 
(circles) 
or NFW profile (filled squares) divided by the detection efficiency for the standard model.
The outside x-axis is the 
redshifts, the outside y-axis is the luminosities. The inner x-axes are the 
distances from the ROSAT PSPC pointing center in pixels. The inner y-axes are 
the difference percentages. \label{fig3}}
\end{figure}

We next show in Figure 2 one effect of modifying the shape of the cluster
emission profile.  We see that it is much easier to detect clusters if their
X-ray emission profile is the  Navarro, Frenk and White (1997; NFW) model and 
that
cooling flows do not have much effect.  Also, at the highest redshifts, but not
shown in a figure here,
clusters which are more elliptical than the average are more easily detected.
These two points demonstrate: (1) that knowledge of shapes of the cluster
emission at high redshift is extremely important for correcting for survey
incompleteness; and, (2) that highly elliptical clusters are more easily detected
than the average, so that this may explain in part why only clusters that appear
to be highly elliptical have been detected at high redshift. For even 
L$_x$ $\sim 10^{45}$ erg cm$^{-2}$ s$^{-1}$ clusters  at $\simeq 0.8$ are
apparently  quite faint.
Keeping the total number of photons emitted constant, but confining the
emission region to a more highly elliptical shape means the cluster has an
intrinsically higher surface brightness and is, hence, easier to detect.
For brevity, we do not show many of the other cases we tested, but the effect 
of the assumed cosmology on the assumed detection efficiency is shown in Figure
3, which is discussed below.

\section{The Simulations Applied To The Bright SHARC Sample}

In Figure 3 we show the results of applying the derived efficiency of detection 
to
the Bright SHARC sample.  There are several comparisons to be made within 
this
figure. First there is the comparison between the standard model result and the
preliminary result of Nichol et al. Here we see the new, refined result 
does not
differ in a statistically significant manner from the Nichol et al. 
work. Second, we see that the effect of assuming different cosmologies is
negligible within the statistical uncertainty. Third we see that the 
statistical uncertainty in the number of clusters in the highest luminosity 
bin is so large because these clusters are so rare and that
a much larger survey than the Bright SHARC is needed 
to
determine if there is any discernible evolution
in the highest luminosity bin.  Fourth, when we compare with local XCLF, we see
there is no statistically significant evidence for evolution of the XCLF as a 
whole and
that these results are qualitatively consistent with the scenario for
hierarchical cluster formation discussed in the introduction.

\begin{figure}[bht]
\vspace*{-.2in}
\hspace{-4.5 in}\centerline{\psfig{file=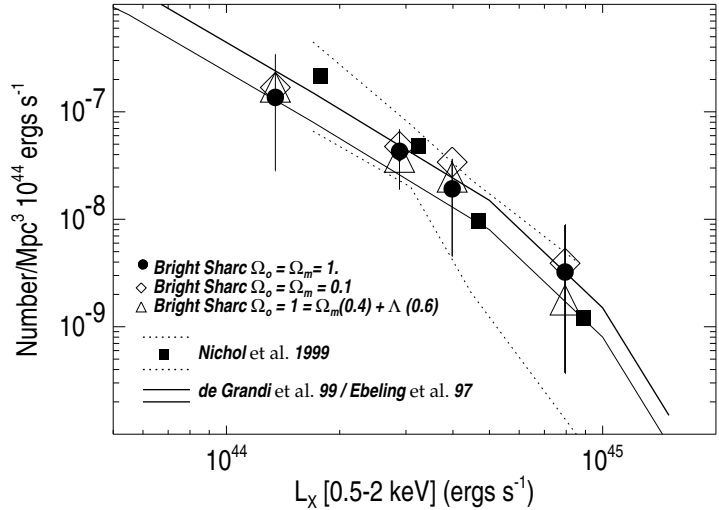,height=3in.} }
\caption{The influence of the cosmological model on the XCLF. The y-axis is the 
number of clusters and the x-axis is the luminosity of the clusters (see Adami
et al 1999). The error bars (vertical lines) are attached to the filled circles.
We plot also the envelope of the local XCLF from Nichol et al (1999): dashed
lines. The 2 solid lines are the local XCLF from de Grandi et al (1999) and 
Ebeling et al (1997). 
\label{fig4}} 
\end{figure}

Follow up work requires both much larger and deeper surveys.

\acknowledgments
We thank the IGRAP meeting organizators and support in part by NASA grant 
NAG5-2432 and by NASA Illinois space grant

\end{document}